\documentclass[a4,12pt,final]{article}
\usepackage{t1enc}
\usepackage[english]{babel}
\usepackage[utf8]{inputenc}
\usepackage{floatflt}
\usepackage{graphicx}
\usepackage{psfrag}
\usepackage{bbm}
\usepackage{amsmath}
\usepackage{amssymb}
\usepackage{showkeys}
\usepackage{ifthen}
\usepackage{subfigure}

\title{Non-Abelian Vortices with a Twist}
\author{Péter Forgács\textsuperscript{1,2}, Árpád Lukács\textsuperscript{1,3} and Fidel A.~Schaposnik\textsuperscript{4}\\
{\small {}\textsuperscript{1} Wigner RCP RMKI, H1525 Budapest, POB 49}\\
{\small {}\textsuperscript{2} LMPT CNRS UMR7350, Universit\'e de Tours, Parc de Grandmont, 37200 Tours, France}\\
{\small {}\textsuperscript{3} Aristotle University, Thessaloniki, Greece}\\
{\small {}\textsuperscript{4} Departamento de Física, Universidad Nacional de La Plata,}\\
{\small Instituto de Física La Plata C.C. 67, 1900 La Plata, Argentina}
}

\def\e{\mathrm{e}}

\def\kihagy#1{}

\def\Tr{\ensuremath{\mathop{\rm Tr}}}

\newcommand{\arxiv}[2][]{
  \ifthenelse{\equal{#1}{}}{
    \href{http://arxiv.org/abs/#2}{\texttt{arXiv:#2}}
  }{
    \href{http://arxiv.org/abs/#2}{\texttt{arXiv:#2 [#1]}}
  }
}

\newcommand{\be}{\begin{equation}}
\newcommand{\ee}{\end{equation}}

\begin{document}
\maketitle

\begin{abstract}
Non-Abelian flux-tube (string) solutions carrying global currents are found
in the bosonic sector of four-dimensional ${\cal N}=2$ super-symmetric gauge theories. The specific model considered here
possesses U(2)$_{\rm local}\times$SU(2)$_{\rm global}$ symmetry, with two scalar doublets in the fundamental representation of SU(2).
We construct string solutions that are stationary and translationally symmetric along the
$x^3$ direction, and they are characterized by a matrix phase between the
two doublets, referred to as ``twist''. Consequently, twisted strings have
nonzero (global) charge, momentum, and in some cases even angular momentum per
unit length.
The planar cross section of a twisted string corresponds to a rotationally
symmetric, charged non-Abelian vortex, satisfying first order Bogomolny-type
equations and
second order Gauss-constraints.
Interestingly, depending on the nature of the matrix phase, some of these solutions even break cylindrical symmetry in $\mathbb{R}^3$.
Although twisted vortices have higher energy than the untwisted ones, they are expected to be linearly stable
since one can maintain their charge (or  twist) fixed with respect to small perturbations.
\end{abstract}

\section{Introduction}\label{sec:intro}
Vortex and string-type solutions appear in many models, and as they have many applications there is an enduring interest for them.  In the plane, a vortex corresponds to a
cross-section of a straight string in three spatial dimensions
in the plane orthogonal to its direction. In the context of spontaneously broken gauge field theories by scalar fields, the paradigm is the Abrikosov-Nielsen-Olesen (ANO)
vortex \cite{ANO} associated to the breaking of an Abelian gauge group. The ANO vortex corresponds to the planar cross-section of a static straight, infinitely long magnetic
flux-tube, with quantized magnetic flux and SO(2) cylindrical symmetry .
ANO vortices have an (integer) winding number, proportional to their quantized magnetic flux, which is also responsible for their stability. For a fixed winding number,
ANO solutions form a one-parameter family, depending on the mass ratio $\beta=m_s/m_v$, where $m_s$, resp.\ $m_v$ denote the mass of the scalar resp.\ of the vector field.
In the special case $\beta=1$, the energy of a vortex is proportional to its winding number \cite{bog,deVegaSchap1}, and surprisingly vortices of like fluxes
do not interact \cite{bog}. For this special value of the coupling, minimal energy vortices satisfy a set of first order - Bogomolny - equations, which are easier
to solve than the field equations \cite{bog,deVegaSchap1}.

In non-Abelian gauge theories with or without a Chern-Simons term, non-Abelian vortices (some of them with an electric charge)
were first obtained in Ref.\ \cite{fidel}.
Since the seminal papers \cite{nav1} started the investigation of vortex-string solutions in supersymmetric non-Abelian gauge theories, the subject continues to
attract attention. A simple model containing the essential features is an $U(N_c)$ gauge theory, coupled to $N_f$ scalar fields in the fundamental representation,
where $N_f\geq N_c$. Vortex solutions in such theories are usually referred to as non-Abelian vortices, (NAVs).
NAVs have attracted considerable interest, since they are at the heart of intriguing relationships between two-dimensional sigma-models and four-dimensional gauge theories. NAVs in a $U(N)$ gauge
theory possess ``orientational moduli'' whose low-energy dynamics is described by  versions of ${\bf CP}^{N-1}$ sigma-models on the string worldsheet.
Moreover, NAVs also held the promise to be relevant to bring us closer to a description of quark confinement; for a review, see \cite{Shifman} and references therein.
Static NAVs are absolute minima of the energy functional in a fixed topological sector, characterized by the winding numbers associated to the Cartan subalgebra of the gauge group.
Moreover NAVs satisfy first order, Bogomolny-type equations admitting rather complicated, static multi-vortex solutions.
Most remarkably a complete description of the NAV moduli space has been found \cite{Auzzi, Eto-review}.
An interesting application of non-Abelian vortices is illustrated by confined monopoles emerging as junctions of NAVs with different moduli \cite{ConfMono, Gorsky}.
The dynamics of NAVs based on the moduli approximation has been worked out in Refs.\ \cite{Collie, Eto-dyonic,Eto-eff}.
In addition to NAVs and related monopoles, domain walls have also received due attention \cite{DW, Shifman},
and a number of intriguing relations between moduli spaces of monopoles and domain walls have been discovered \cite{HT}.

In this paper we point out that by allowing for space-time-dependent phases among the scalar fields, which we shall call ``twisting'', new families of charged vortex-strings arise. We shall restrict our attention to a U(2) gauge group broken by two scalar doublets with an appropriate scalar potential compatible with super-symmetry (SUSY), but a
generalization to other groups should not be too difficult. It is known, that imposing the usual space-time symmetries on field configurations leading to vortex-type solutions, such as translational invariance in
time and along the, say, $x^3$ or $z$ direction, scalar fields may have a phase, with a linear dependence on $(t\,,z)$ \cite{Abraham, FRV,GV}.
In general, the energy of {\it stationary} configurations with a nontrivial $(t\,,z)$-dependent phase is bigger than that of the static ones, however this does not make them necessarily unstable \cite{Abraham}.

In this paper we systematically
investigate straight vortex-string solutions in the simplest theory admitting non-Abelian vortices, when the scalar fields, $\Phi$, possess a $(t\,,z)$-dependent phase, i.e.\
$$
\Phi(x^\nu) = \Phi(x^i)e^{\frac{i}{2}M\omega_\alpha x^\alpha}\,,\quad i=1,2,\ \alpha=0,3\,,
$$
where the flavours correspond to the columns of the ($2\times2$) matrix $\Phi$, $M$ is a constant Hermitian matrix acting on the flavour indices, and $\omega_\alpha$ is vector in $(t\,,z)$ plane.
As it is well known \cite{Abraham, FRV, Ferreira} when $\omega_\alpha$ is a {\it light-like vector}, the field equations in the $(x^1\,,x^2)$ plane decouple completely from those in the $(t\,,z)$ directions.
The chromo-electric components in the $(t\,,z)$ plane are determined by a set of linear, second order equations, (Gauss constraints), depending on the solution in the $(x^1\,,x^2)$ plane.
Simple analysis of the Gauss constraints makes it very plausible that any NAV can be twisted by an arbitrary twisting matrix, although we have not attempted to formally prove this.
There are global currents flowing in the $z$ direction of an $M$-twisted vortex-string which has nonzero charge,
momentum and, unless $M$ is specially aligned in internal space,
angular momentum.
A twisted NAV has some genuine three-dimensional structure, as there are global currents flowing along its axis,
the fields have some nontrivial components orthogonal to its symmetry plane,
hence it is more conveniently thought of as a twisted string.

The simplest twisted NAV can be obtained from the rotationally symmetric ``elementary'' vortex solution of Ref.\
\cite{nav1}, in which case, for a general
twisting matrix the problem reduces to a single second-order Gauss-constraint, which can be easily analyzed.
In fact our twisted NAV solution turns out to be the same object as the ``dyonic'' vortex solution found in a ''mass-deformed'' SUSY gauge-field theory \cite{Eto-dyonic}.
The bulk of our paper concerns the twisting of ``composite-coincident'' NAVs \cite{Auzzi}, corresponding to the rotationally symmetric superimposition of elementary vortices. We characterize these
superimposed NAVs by a relative winding number and a moduli parameter.
Interestingly we find that when the twisting matrix $M$, contains off-diagonal components the twisted string looses cylindrical symmetry in three dimensions.

The energy of an $M$-twisted string is higher than that of an untwisted one, it is given as a sum of
the usual ``magnetic'' energy per unit length proportional to the magnetic flux, and of an ``electric'' contribution due
to the rotating phase.
The actual magnitude of the ``electric'' contribution depends essentially on the components of the twisting matrix, $M$.
When $M$ is diagonal the magnitude of the electric energy of a twisted NAV is typically much smaller than its magnetic
one, in fact, it can be made arbitrarily small when the untwisted NAV is sufficiently close
to a diagonal one. In the case when $M$ contains off-diagonal components, the magnitude of the electric energy is comparable to the magnetic one.

It seems to us that all the above-mentioned properties make twisted NA vortex-strings of some interest and worthy of further investigations.

The plan of the paper is the following.
We introduce in Sec.\ \ref{sec:twist} the theory that we shall study and proceed to the dimensional reduction by splitting
four-dimensional Minkowski space into planar and temporal-longitudinal coordinates. Making an appropriate Ansatz we get a
dimensionally reduced Lagrangian. Assuming that coupling constants satisfy Bogomolny conditions we present the minimal energy
first-order equations satisfied by untwisted solutions and write the energy as a sum of electric and magnetic contributions.
We discuss how the rotational symmetry of the Ansatz, or the absence thereof, depends on the properties of the twisting.
In Sec.\ \ref{sec:twistedv} we consider twisted vortices, both elementary (Sec.\ \ref{sec:elemv}) and composite (Sec.\ \ref{sec:compv})
ones, presenting the numerical study of the solutions. In Section \ref{sec:sum} we summarize the results and present our conclusions. Finally our notations and conventions are given in Appendix \ref{app:conventions}, and some numerical data is given in
Appendix \ref{app:numdata}.

\section{Dimensional reduction}\label{sec:twist}
The  3+1-dimensional theory we consider is defined by the following Lagrangian:
\begin{equation}
  \label{eq:lag_3d}
  \mathcal{L} = -\frac{1}{4g_1^2} F_{\mu\nu}F^{\mu\nu} -\frac{1}{4g_2^2}G_{\mu\nu}^a G^{\mu\nu a} +
   \Tr(D_\mu \Phi)^\dagger D^\mu\Phi -V\,,
\end{equation}
where $F_{\mu\nu}$ is the Abelian field strength tensor and $G_{\mu\nu}^a$ $(a=1,2,3)$ is the non-Abelian,
$SU(2)$ one. The two scalar doublets are encoded in the matrix
$(\Phi)_{iA}$ with $i=1\,,2$ being the gauge (``colour'') and $A=1\,,2$ the ``flavour'' index; the trace is taken over
the flavour indices; the scalar potential, $V=V_1+V_2$, can be written more explicitly as
\begin{align}
  \label{eq:pot}
  V_1 &= \frac{\lambda_1}{8}(\Tr \Phi^\dagger \Phi - 2 \xi)^2
  =\frac{\lambda_1}{8}\left(\Phi_A^\dagger\Phi_A-2\xi\right)^2\,,\\
V_2 &=\frac{\lambda_2}{8}(\Tr\Phi^\dagger \sigma^a \Phi)^2=
\frac{\lambda_2}{8}(\Phi_{Ai}^\star\sigma^a_{ij}\Phi_{jA})^2=
\frac{\lambda_2}{8}\left[(\Phi_1^\dagger\Phi_1-\Phi_2^\dagger\Phi_2)^2
+ 4|\Phi_2^\dagger\Phi_1|^2\right]\,.
\end{align}
where summation over repeated indices is understood, except when the contrary is indicated. For more details on
notations and conventions, see Appendix \ref{app:conventions}.

The fields transform under the $U(2)$ gauge symmetry as
\begin{equation}
  \label{eq:gaugeact2}
  \Phi \to \exp(i\Lambda)U\Phi\,,
  \quad C_\mu \to U C_\mu U^\dagger +
  2iU\partial_\mu U^\dagger\,,\quad A_\mu \to A_\mu +2\partial_\mu \Lambda\,,
\end{equation}
where $U(x)\in SU(2)$, $\Lambda(x)$ is a real function. The flavor symmetry acts from the right on the scalars as
\begin{equation}
  \label{eq:flavoract}
  \Phi \to \Phi V\,,\quad V\in SU(2)\,.
\end{equation}

Let us now consider stationary and translationally symmetric fields in the $x^3$ direction.
It will turn out to be convenient to split four-dimensional Minkowski coordinates
as $x^\mu=(x^\alpha\,,x^i)$ with $\alpha=0,3$, $i=1,2$.
Since the symmetries are generated by two commuting vector fields, there exists a gauge where the symmetric gauge fields are
simply independent of the coordinates $x^\alpha$ \cite{FM}. At the same time, however, the scalar doublets, $\Phi$, may still depend
linearly on $x^\alpha$ through an  SU(2) phase,  i.e.\
the most general symmetric Ansatz can be written as
\begin{subequations}
  \label{eq:FMred}
  \begin{align}
   A_\mu(x^\nu) &= (A_i(x^j), A_\alpha(x^j))\,,\label{ansA}\\
  C_\mu^a(x^\nu) &= (C_i^a(x^j), C_\alpha^a(x^j))\,,\label{ansC}\\
  \Phi(x^\nu) &= \Phi(x^i)e^{\frac{i}{2}M\omega_\alpha x^\alpha}\,,
  \label{ansPhi}
   \end{align}
\end{subequations}
where $M$ is a constant Hermitian matrix,
\be\label{eq:MPauli}
M=m^{\hat{a}}\sigma^{\hat{a}} = m^0{\bf 1} + m^a \sigma^a,
\ee
$\omega_\alpha$ is a vector in the $(t\,,z)$ plane.
Straight flux-tube/string solutions of the theory defined by Eq.\
\eqref{eq:lag_3d}, described by the Ansatz \eqref{eq:FMred} with a non-trivial $M$, will be referred to as `twisted' strings.

The form of the Ansatz, Eq.\ (\ref{eq:FMred}) also restricts the symmetries of the model. Those symmetries, which preserve
the Ansatz are flavour transformations which commute with the twist matrix $M$,
\begin{equation}
  \label{eq:FlavorAns}
  \Phi \to \Phi V\,,\quad V=\exp(iM\delta)\,,
\end{equation}
and all gauge transformations where $U\partial_\alpha U^\dagger$ and $\partial_\alpha \Lambda$ only depend on $x^i$ and not on $x^\alpha$.
The flavor-current generating the transformations \eqref{eq:FlavorAns} can be written as
\be\label{eq:Qcurr}
K_\mu = m^{\hat{a}} K^{\hat{a}}_\mu\,,
\ee
with $K^{\hat{a}}_\mu$  defined in Eq.\ \eqref{eq:flcurr}.

Being Hermitian, the twist matrix, $M$, can always be diagonalized by a unitary matrix, $V_M$, such that
$M=V_M M_D {V_M}^\dagger$, with $M_D$ being diagonal, i.e.\
\begin{equation}
  \label{eq:DiagM}
  \Phi(x^i)e^{\frac{i}{2}M\omega_\alpha x^\alpha}=\Phi(x^i) V_M e^{\frac{i}{2}M_D\omega_\alpha x^\alpha} {V_M}^\dagger\,.
\end{equation}
Therefore, twisting a configuration with the matrix $M$, is equivalent to
twisting a suitably flavour-transformed (with $V_M$) configuration with the diagonal matrix, $M_D$.
As the theory considered is SU(2) flavour symmetric it makes no difference which form of twisting one uses.
In what follows we shall work with the non-transformed twisting matrix, $M$ in Eq.\ \eqref{ansPhi}.

The Ansatz (\ref{eq:FMred}) yields the dimensionally reduced Lagrangian
\begin{equation}
  \label{eq:LagFM1}
  \begin{aligned}
  \mathcal{L} = &-\frac{1}{4g_1^2}F_{ij}^2 + \frac{1}{2g_1^2}\partial_iA_\alpha\partial_iA^\alpha-
  \frac{1}{4g_2^2}(G_{ij}^a)^2+\frac{1}{2g_2^2} D_iC_\alpha^a D_iC^{\alpha a}\\
      &-\frac{1}{4g_2^{2}}[({C}^a_\alpha{C}^{a\alpha})^2-(C^a_\alpha{C}^a_\beta)(C^{b\alpha}{C}^{b\beta})]
      \\
      &-\Tr(D_i\Phi)^\dagger D_i\Phi + \frac{1}{4}\Tr (\omega_\alpha \Phi M -
      \mathbb{C}_\alpha\Phi)^\dagger (\omega^\alpha \Phi M - \mathbb{C}^\alpha\Phi) - V\,,
  \end{aligned}
\end{equation}
where $\mathbb{C}_\alpha = A_\alpha + C_\alpha^a \sigma^a$. It is now convenient to introduce basis vectors in the
$x^\alpha$ plane, ($\omega_\alpha\,,\bar{\omega}^\alpha$) such that $\omega_\alpha {\bar{\omega}}^\alpha =0$ if
$\omega^2= \omega_\alpha \omega^\alpha\ne 0$, while for the light-like case ($\omega^2 = {\bar{\omega}}^2=0$)
$\omega_\alpha {\bar{\omega}}^\alpha \ne0$. We remark that in Eq.\ \eqref{eq:LagFM1} only the $\alpha$ indices
are raised or lowered by the induced Minkowskian metric in the $(x^0,x^3)$ plane, and the repeated $i\,,j$-type lower indices are
summed with $(+,+)$ signature.
It turns out that the dimensionally reduced Lagrangian \eqref{eq:LagFM1}, is closely related to the trivial reduction from
four to two dimensions of the ``mass deformed'' theory considered in Ref.\ \cite{Eto-dyonic}.
In the absence of twist, $M\equiv0$, \eqref{eq:LagFM1} corresponds to the trivial reduction of the theory to two dimensions,
whose solutions are the non-Abelian vortices discussed in detail in Ref.\ \cite{Eto-review}.
With respect to the $\omega\,,\bar{\omega}$ basis the gauge field components in the $(x^0,x^3)$ plane are expressed as
\begin{equation}
  \label{eq:basis2}
  A_\alpha = \omega_\alpha A + {\bar{\omega}}_\alpha \bar{A}\,,\quad C_\alpha^a = \omega_\alpha C^a +
  {\bar{\omega}}_\alpha {\bar{C}}^a\,.
\end{equation}
The field equations can be grouped according to variations with respect to $\bar{A}\,,\bar{C}$ resp. $A\,,C$.
Consider first the variational equations with respect to $\bar{A}\,,\bar{C}$:
\begin{subequations}
\label{homogenous_eqs}
\begin{align}
\triangle \bar{A}&=
\frac{g_1^2}{2}\,\Tr(\Phi^\dagger\bar{\mathbb{C}}\Phi)\,,\label{hom1}\\
{\hat D}_i {\hat D}_i\,\bar{C}^a &=
\frac{g_2^2}{2}\,\Phi_A^\dagger(\bar{A}\sigma^a+\bar{C}^a)\Phi_A-\Delta^a\,,\label{hom2}\\
\Delta^a&=\begin{cases}\omega^2\,C^b(C^b\bar{C}^a-\bar{C}^bC^a)&\mbox{if}\ \omega^2\ne0\\
                        2(\omega\bar{\omega})\,\bar{C}^b(C^b\bar{C}^a-\bar{C}^bC^a)&\mbox{if}\ \omega^2=\bar{\omega}^2=0\,,
                        \end{cases}
\end{align}
\end{subequations}
where $\triangle=\partial_i\partial_i$ with Euclidean metric summation.
From Eqs. \eqref{homogenous_eqs} we obtain the following integral identity:
\be\label{Gauss_id}
\int d^2x\left[
\frac{1}{2}{\mathbf{\bar\Delta}}
-
\frac{1}{g_1^2}(\partial_i\bar{A})^2-\frac{1}{g_2^2}(\partial_i\bar{C}^a)^2
-\frac{1}{2}\Phi_A^\dagger\bar{\mathbb{C}}^2\Phi_A+\frac{1}{g_2^2}\bar{C}^a\Delta^a
\right]=0\,,
\ee
where ${\mathbf{\bar\Delta}}=\triangle\left({\bar{A}}^2/g_1^2+{\bar{C}}^a{\bar{C}}^a/g_2^2\right)$.
Assuming finite-energy boundary conditions and global regularity, the integral of ${\mathbf{\bar\Delta}}$ is zero.
For the case when $\omega$ is light-like, $\omega^2=0$, one finds $\bar{C}^a\Delta^a=0$, therefore Eq.\ \eqref{Gauss_id}
enforces $\bar{A}\equiv0$, $\bar{C}\equiv0$.
When $\omega^2<0$, Eq.\ \eqref{Gauss_id} implies once more the vanishing of $\bar{A}$ and
$\bar{C}$, since then $\bar{C}^a\Delta^a\leq0$.
In the case of a space-like $\omega$ vector, $\omega^2>0$, Eq.\ \eqref{Gauss_id} is not sufficient to exclude
the existence of non-trivial solutions of the Gauss constraints, \eqref{homogenous_eqs}.
It is consistent, however, to assume $\bar{A}\equiv0$, $\bar{C}\equiv0$, even for $\omega^2>0$, since $\bar{A}$, $\bar{C}$
satisfy homogenous equations.
Assuming $\bar{A}\equiv0$, $\bar{C}\equiv0$ the remaining field equations are
\begin{subequations} \label{eq:FeqFM}
\begin{align}
\frac{1}{g_1^2} \triangle A &= \frac{1}{2}\Tr\left[ (\Phi^\dagger\mathbb{C}-M\Phi^\dagger)\Phi \right]\,,\label{req1}\\
\frac{1}{g_2^2} \hat{D}_i \hat{D}_i C^a &= \frac{1}{2}\Tr\left[\left(C^a\Phi^\dagger+(A-M)\Phi^\dagger\sigma^a\right)\Phi\right]\,,\label{req2}\\
\frac{1}{g_1^2}\partial_i F_{ij} &= \frac{i}{2}\Tr\left[ \Phi^\dagger D_j \Phi - D_j\Phi^\dagger \Phi\right]\,,\label{req3}\\
\frac{1}{g_2^2}\hat{D}_i G_{ij}^a &= \frac{i}{2}\Tr\left[ \Phi^\dagger \sigma^a D_j \Phi - D_j\Phi^\dagger \sigma^a\Phi\right]
 -\frac{\omega^2}{g_2^2}\varepsilon^{abc}C^b D_jC^c\,,\label{req4}\\
D_iD_i\Phi &= \frac{\partial V}{\partial \Phi^\dagger} - \frac{\omega^2}{4}\left[ (\Phi M-2\mathbb{C}\Phi)M +
\mathbb{C}^2\Phi\right]\,. \label{req5}
\end{align}
\end{subequations}
The total energy of an $M$-twisted string can be written as the sum of an ``electric'' and of a ``magnetic'' part:
\be
E=\int\, d^2x T^0_{~0}=\int({\cal E}_0+{\cal E}_1)\, d^2x\,
\equiv E_0+E_1 .
\ee
where the ``electric'', resp.\ ``magnetic'' densities, ${\cal E}_0\,,{\cal E}_1$ are defined as
\begin{align} \label{En1}
{\cal E}_0&=
\frac{1}{2g_2^2}\,G^a_{\alpha i}G^a_{\alpha i}
+\frac{1}{2g_1^2}F_{\alpha i}F_{\alpha i}
+(D_\alpha\Phi_A)^\dagger D_\alpha\Phi_A \\
&=(\omega_0^2+\omega_3^2)\left[\frac{1}{2g_1^2} (\partial_i A)^2 +
 \frac{1}{2g_2^2}(D_i C^a)^2 + \frac{1}{4}\Tr (\Phi M - \mathbb{C}\Phi)^\dagger (\Phi M - \mathbb{C}\Phi)  \right]\,,  \nonumber\\
{\cal E}_1&=\frac{1}{4g_2^2}\,(G^a_{ik})^2
+\frac{1}{4g_1^2}(F_{ik})^2+
|D_i\Phi_A|^2+V\,.
\end{align}
A straightforward computation shows that using the Gauss constraints, Eqs.\ \eqref{req1}--\eqref{req2},
the ``electric'' density, ${\cal E}_0$
can be expressed as
\be\label{E0}
{\cal E}_0=\frac{1}{4}(\omega_0^2+\omega_3^2)({\mathbf{\Delta}}+Q)\,,
\ee
where ${\mathbf{\Delta}}=\triangle\left(A^2/g_1^2+C^aC^a/g_2^2\right)$ and
\be\label{Q}
Q= \Tr \left[ \Phi^\dagger(\Phi M- \mathbb{C}\Phi) M \right]\,.
\ee
Note that the $(t,z)$-components of the flavor current in Eq.\ \eqref{eq:Qcurr} can also be expressed
in terms of $\omega$ and $Q$ as
\be\label{flavorcurr}
K_\alpha = -\frac{\omega_\alpha}{2}Q\,.
\ee
In most work on non-Abelian vortices the supersymmetry-induced relations between the couplings,
 $\lambda_1=g_1^2$, $\lambda_2=g_2^2$, have been assumed. These relations ensure that ${\cal E}_1$ can be expressed as a sum of squares
 and a topological term, leading to  first-order, Bogomolny-type equations in the $(x^1,x^2)$ plane.
Then minimal-energy, {\sl untwisted} solutions of the second-order field equations \eqref{req3}--\eqref{req5} are obtained
by solving the following first-order equations:
\begin{subequations} \label{eq:Bog}
   \begin{align}
    F_{ik} &= \mp \frac{g_1^2}{2}\epsilon_{ik}(\Tr\Phi^\dagger\Phi -2\xi)\,,\label{F-1st}\\
    G_{ik}^a &= \mp \frac{g_2^2}{2}\epsilon_{ik}\Tr\Phi^\dagger \sigma^a\Phi\,,\label{G-1st}\\
    D_i\Phi &= \mp i\epsilon_{ik}D_k\Phi\,,\label{Phi-1st}
  \end{align}
\end{subequations}
while for solutions of Eqs.\ \eqref{eq:Bog} the ``magnetic'' energy density simplifies to
\be\label{E1}
{\cal E}_1=\pm\frac{\xi}{2}\epsilon_{ik}F_{ik}\mp i\epsilon_{ik}\partial_i\left(\Phi_A^\dagger D_k\Phi_A\right)\,.
\ee
Equation \eqref{E1} implies that the total ``magnetic'' energy, $E_1$, is given by the net Abelian flux through the $(x^1,x^2)$ plane.
Let us quote here the actual value of the ``magnetic energy'' in Eq.\ \eqref{E1}, $E_1$, for configurations considered
in this paper, characterized by winding numbers $n_A\,,m_A$ as in Eq.\ \eqref{gen-winding} and subject to
the first-order Eqs.\ \eqref{eq:Bog}:
\be\label{magnetic-en}
E_1=2\pi\xi|n_1+m_2|=2\pi\xi|n_1+m_1+N|\,,
\ee
which of course also holds for the Ansatz \eqref{minAnsatz}.
In the case when $\omega_\alpha$ is light-like, the field equations in the $(x^1,x^2)$ plane,
Eqs.\ \eqref{req3}--\eqref{req5},
decouple from the Gauss-type constraints, Eqs.\ \eqref{req1}--\eqref{req2}, and become identical to those corresponding
to untwisted vortices.
Therefore the problem of finding twisted non-Abelian strings for $\omega^2=0$ reduces to solving Eqs.\ \eqref{req1}--\eqref{req2}
in the {\sl background} of a non-Abelian vortex in the $(x^1,x^2)$ plane. In the present paper we shall consider a light-like
$\omega$ vector,
and concentrate on twisting vortex solutions of minimal energy satisfying the first-order Eqs.\ \eqref{eq:Bog}.
It is left for future work to clarify if for $\omega^2\ne0$
there exists solutions analogous to the twisted vortices of Refs.\ \cite{FRV,GV}.

An $M$-twisted string has momentum flowing along the $z$ direction, which is easily obtained from the stress-energy tensor.
The longitudinal momentum, $P= T_{03}$, carried by a twisted string, can be recast exploiting the Gauss constraints,
Eqs.\ \eqref{req1}--\eqref{req2}, as
\begin{equation}
  \label{eq:T03FM}
 P=T_{03} = \frac{1}{2}\omega_0 \omega_3 ({\mathbf{\Delta}}+Q)\,.
\end{equation}
$M$-twisted strings may also have angular momentum, $J$, as it can be seen from the angular momentum density
\begin{equation}
  \label{eq:T02FM}
 J= T_{0\vartheta} = \frac{1}{g_1^2} F_{0r} F_{\vartheta r} + \frac{1}{g_2^2}G_{0 r}^a G_{\vartheta r}^a +
  \Tr D_0\Phi^\dagger D_\vartheta\Phi + \Tr D_\vartheta\Phi^\dagger D_0\Phi\,;
\end{equation}
however, to compute $J$ in a more explicit form, one needs a parametrization of the angle dependence of the fields.
This will be presented in the next section.

\subsection{The Ansatz; rotational symmetry and its loss}\label{sec:minAns}
Let us now impose rotational symmetry in the $x^1,x^2$ plane to
the fields. Denoting the usual polar coordinates
in the  plane as  $x^1=r\cos\vartheta$,  $x^2=r\sin\vartheta$,
rotational symmetry implies that by $\vartheta$-dependent gauge
transformations one can always achieve
\be\label{gen-winding}
\partial_\vartheta \{A_\mu,C_\mu^a\}=0\,,\quad
\Phi_A(x^i)=\left(\exp({in_A\vartheta})\Phi_{A\,1}(r),\exp({im_A\vartheta})\Phi_{A\,2}(r)\right)\,,\;\;A=1,2\,.
\ee
In order to ensure consistency with the U(2) gauge and the global  SU(2) flavour symmetry, the integers,
$n_A,m_A$, satisfy the following relation:
$n_2-n_1=m_2-m_1=N$, which can also be expressed on the two scalar doublets as
\be\label{rel-winding}
\Phi(x^i) =
\begin{pmatrix} \phi_1(r)e^{in_1\vartheta} & \psi_1(r)e^{in_1\vartheta} \\
\phi_2(r)e^{im_1\vartheta} & \psi_2(r) e^{im_1\vartheta} \end{pmatrix}
\begin{pmatrix}
 1&0\\0&e^{iN\vartheta}
\end{pmatrix} = \Phi_0(x^i)e^{i{\bf N}\vartheta}
\,,
\ee
where ${\bf N}={\rm Diag}\{0,N\}$, which encodes the relative winding between the two flavours. Since in general
the twisting
matrix, $M$, mixes the two flavours, when $N\ne0$ in Eq.\ \eqref{rel-winding}
one can immediately see, that the right hand sides of Eqs.\ \eqref{req1}--\eqref{req2}
depend explicitly on
$\vartheta$, breaking rotational symmetry in the $x^{\alpha}$ direction, i.e.
\be
\partial_\vartheta \{A_\alpha,C_\alpha^a\}\ne0\,.
\ee
It is easy to see that the condition to ensure rotational symmetry of solutions of the field equations
\eqref{eq:FeqFM} can be written as
\begin{equation}
  \label{eq:symmcond}
  [ M,{\bf N} ] = 0\,,
\end{equation}
implying that the twisting matrix is diagonal.
This happens whenever $M$ does not contain terms proportional to $\sigma^1\,,\sigma^2$.
Obviously, the anisotropy generated by twisting matrices not commuting with ${\bf N}$,
would simply rule out the possibility to consider rotationally symmetric configurations.
Remarkably in the case of a light-like
twist vector, due to the decoupling of the field equations in the $(x^1,x^2)$ plane from the
equations with $x^{\alpha}=(x^0,x^3)$ components, \eqref{req1}--\eqref{req2}.
This decoupling allows for solutions which actually break rotational symmetry in the $x^{\alpha}$ direction.
Whenever $M$ does not commute with ${\bf N}$ the corresponding
$M$-twisted strings have rotationally symmetric spatial sections
in any plane orthogonal to the $x^3$-axis; however, the complete configuration is not
rotationally symmetric in the whole space-time.

Keeping the possibility of breaking rotational symmetry in mind, we now present our Ansatz.
By {\sl singular} U(2) gauge transformations (linear in $\vartheta$) on the scalars in Eq.\ \eqref{rel-winding}
we can achieve
$n_1=m_1=0$. Furthermore, by assuming that the functions, $\Phi_{A\,i}(r)$, are all {\sl real}, one reduces
the number of
scalar fields from eight to four  (minimality of the Ansatz).
Then from Eq.\ \eqref{G-1st} it follows that $C_\vartheta^2={\rm const}$, which can be set to zero.

Finally choosing the radial gauge, our Ansatz can be written as
\begin{subequations}\label{minAnsatz}
\begin{align}
A_\alpha&=A(r,\vartheta)\omega_\alpha\,,\;\;\;\; A_r=0\,,\;\;\;\;A_\vartheta=a(r)\,,\\
C_\alpha^a&=C^a(r,\vartheta)\omega_\alpha\,,\;\;\;C^a_r=0\,,\;\;\;\;
C^a_\vartheta=\left\{c_1(r),0\,,c_3(r)\right\}\,,
\Phi(x^i)&=
\begin{pmatrix} \phi_1 (r) & \psi_1(r)\e^{iN\vartheta} \\ \phi_2(r) & \psi_2(r) \e^{iN\vartheta}\end{pmatrix}\,,
\end{align}
\end{subequations}

We can now display the angular momentum, $J$ for our Ansatz \eqref{minAnsatz} in a more explicit form.
Exploiting Eqs.\ \eqref{req1}--\eqref{req2} we obtain
\begin{equation}
  \label{eq:T02FMred}
  J = \omega_0 \left( \tilde{\bf\Delta} + \tilde{Q} \right)\,,\quad
\tilde{Q} = \Tr \left[\Phi^\dagger\Phi({\bf N}M+M{\bf N})-2\Phi^\dagger \mathbb{C}\Phi {\bf N} \right]
\end{equation}
where $\tilde{\bf\Delta} = \partial_i(a\partial_i A)/g_1^2 + \partial_i(C^a_\vartheta\hat{D}_iC^a)/g_2^2$.
It is worthwhile to point out, that $\tilde{Q}$ is a combination of the flavour charge densities [see Eq.\ (\ref{eq:flcurr})],
$\tilde{Q} = -4 n^{\hat{a}}K_0^{\hat{a}}/\omega_0$, with the coefficients
$n^a = \Tr(N\sigma^a)/2$, $n^0 = \Tr N/2$.

Let us now write out the explicit form of the first-order equations \eqref{eq:Bog} for our Ansatz:
\begin{subequations}
\begin{align}\label{eq:Bogeqs2}
a'\pm \frac{g_1^2}{2}r\left(\phi_1^2+\phi_2^2+\psi_1^2+\psi_2^2-2\right)&=0\,,\\
              c_3'\pm \frac{g_2^2}{2}r(\phi_1^2-\phi_2^2+\psi_1^2-\psi_2^2)&=0\,,\\
                            r\phi_1'\pm\frac{1}{2}[(a+c_3)\phi_1+c_1\phi_2]&=0\,,\\
                            r\psi_2'\pm\frac{1}{2}[(a-c_3-2N)\psi_2+c_1\psi_1]&=0\,,\\
                        c_1'\pm g_2^2r(\phi_1\phi_2+\psi_1\psi_2)&=0\,,\\
                         r\phi_2'\pm\frac{1}{2}[(a-c_3)\phi_2+c_1\phi_1]&=0\,,\\
                         r\psi_1'\pm\frac{1}{2}[(a+c_3-2N)\psi_1+c_1\psi_2]&=0\,,
\end{align}
\end{subequations}
where for convenience we have chosen units such that $\xi=1$.
The vacuum manifold of ${\cal E}_1$ for the Ansatz \eqref{minAnsatz} corresponds to the fix-point manifold of Eqs.\
(\ref{eq:Bogeqs2}-g), a curve, which can be parametrized as
\begin{alignat}{4}\label{red_vac}\nonumber
\phi_1&=\cos\alpha\,,\quad &\phi_2&=\sin\alpha\,,\ &\psi_1&=-\sin\alpha\,,\ &&\psi_2=\cos\alpha\,,\\
a&=N\,,\ &c_1&=-N\sin(2\alpha)\,,\ &c_3&=-N\cos(2\alpha)\,,\quad  &&0\leq\alpha\leq 2\pi\,.
\end{alignat}

\section{Twisted vortices}\label{sec:twistedv}

\subsection{Twisted elementary vortices}\label{sec:elemv}
The simplest non-Abelian vortex solution is a `diagonal' one, with just
$\phi_1$, $\psi_2$, $a$, $c_3$ being non-trivial and subject to Eqs.\ (\ref{eq:Bogeqs2}-g), while $\phi_2$, $\psi_1$,
 $c_1$ are all
zero. In this case the `vacuum angle', $\alpha=0$. Such solutions have been thoroughly investigated in
Ref.\ \cite{Eto-review}.
A larger family of `elementary' vortex solutions of Eqs.\ (\ref{eq:Bogeqs2}-g), with $\phi_2$, $\psi_1$, $c_1$
being also nontrivial,
can be obtained by a `colour-flavour' transformation from a ``diagonal'' one  \cite{Eto-review}.
We note that for non-diagonal NAVs the parameter $\alpha$ is different from zero.
For diagonal vortices the relative winding between the two doublets is always trivial, $N=0$, which remains so for
the general elementary vortices. The general form of the elementary solution can be written as:
\begin{equation}
  \label{eq:elemV}
  \Phi = \phi_+ \mathbbm{1} + \phi_- n^a \sigma^a\,,\quad A_{\vartheta} =a(r)\,,\quad C^a_{\vartheta} = n^a c_3\,,
  \quad \phi_\pm =   (\phi_1\pm \psi_2)/2\,,
\end{equation}
where $n^a$ is the `orientational' unit vector of an
elementary NAV, which can be parametrized by the two spherical angles as
$n^a=(\sin\alpha\cos\beta\,,\sin\alpha\sin\beta\,,\cos\alpha)$.
In this case the two moduli parameters are just the angles, $\alpha$ and $\beta$ \cite{Shifman,Eto-review}.

We note here that for the most general (with the two moduli) elementary vortex solution in Eq.\ \eqref{eq:elemV}
$C^2_{\vartheta}\ne0$, and hence for $\beta\ne0$ it is not in the form of the minimal Ansatz \eqref{minAnsatz};
this has no influence, however, on our twisting of this solution.
It is now simple to generalize, or deform the elementary vortex solution by $M$.
Parametrizing the twisting matrix as in Eq.\ (\ref{eq:MPauli}),
a short computation shows that the Ansatz
\begin{equation}
  \label{eq:elemVSigma}
  A=m^0\,,\quad C^a = (m\!\cdot\! n)[1-C(r)]n^a + C(r)m^a \,,\quad m\!\cdot\! n=m^a n^a\,,
\end{equation}
containing just a single radial function, $C(r)$,
reduces the Gauss constraints, Eqs.\ \eqref{req1}--\eqref{req2}, to a single second-order inhomogeneous equation
for $C(r)$:
\begin{equation}
  \label{C-eq}
  \frac{1}{r}(r{{C}}')' -\frac{c_3^2}{r^2}C = g_2^2\left[ \phi_{+}^2 (C-1) + \phi_{-}^2(C+1)\right]\,.
\end{equation}
Without going into more rigorous mathematics one can easily convince oneself that Eq.\ \eqref{C-eq} admits
 a unique solution subject to boundary conditions guaranteeing regularity at $r=0$ and at $r\to\infty$. Conversely, using
 maximum-principle-type arguments it is not difficult to show that all globally regular solutions are necessarily of
 the form of the Ansatz in Eq.\ \eqref{eq:elemVSigma}.

The charge density $Q$, determining the energy and momentum, of the elementary string, is given as:
\be
 \label{eq:elemVQ}
  Q = \left(m\!\cdot\!m - (n\!\cdot\!m)^2\right) \left[\phi_1^2+ \psi_2^2-2 C(r)\phi_1\psi_2\right]\,,
\ee
As one can easily see from Eq.\ \eqref{eq:T02FMred} the angular momentum density of twisted elementary vortices vanishes.
As already pointed out, this subsection reproduces and extends
previously obtained results of Ref.\ \cite{Eto-dyonic} using different methods,
although our starting was point rather different.

\subsection{Twisted coincident composite vortices}\label{sec:compv}
Next we consider a vortex solution with nonzero relative winding between the two flavours. Assuming the minimal Ansatz,
 the vortex in the $(x^1\,,x^2)$ plane is rotationally symmetric, satisfying the first-order equations 
(\ref{eq:Bogeqs2}-g).
Such solutions have already been analyzed in Ref.\ \cite{Auzzi}, where it has also been pointed out that they correspond to
superimposed vortices on top of each other. Therefore such vortices can be considered as composed of elementary ones.
As we have already argued, deforming composite vortices with a general matrix, $M$ induces a nontrivial
angle dependence in the $(x^0\,,x^3)$ plane.

Without losing generality one can parametrize the twisting matrix as
\be\label{M-parameters}
M=\frac{s}{2}\left({\mathbf{1}}-\sigma^3\right)+\frac{m}{2}\left(\cos(\mu)\sigma^1+\sin(\mu)\sigma^2\right)\,.
\ee
Taking into account the possible angle dependence of the $\omega_\alpha$ components of the gauge fields, we introduce the
following decomposition:
\be\label{Fourier}
A=sA_0+mA_{+}e^{i(N\vartheta+\mu)}+mA_{-}e^{-i(N\vartheta+\mu)}\,,\quad C^a=sC^a_0+mC^a_{+}e^{i(N\vartheta+\mu)}
+mC^a_{-}e^{-i(N\vartheta+\mu)}\,,
\ee
together with the conditions $A_{+}={A^{*}_{-}}$, $C^a_{+}=({C^a_{-}})^*$ ensuring the reality of the fields; moreover
$A_0,A_\pm$, and $C^a_0,C^a_\pm$ are functions of $r$.
Then the corresponding Gauss constraints can be put in the form
\begin{subequations}
\begin{align}\label{Gauss-A0}
\triangle_{r} A_0&=g_1^2(\eta A_0+\eta^{\bar a} C^{\bar a}_0-\frac{1}{2}\eta^0_2)\,,\\\label{Gauss-C0}
\triangle_r C^{\bar a}_0+\frac{\epsilon^{\bar{a}\bar{b}}c_{\bar b}}{r^2}C^{\scriptscriptstyle(13)}_0 &=
g_2^2(\eta C^{\bar a}_0 + \eta^{\bar a} A_0-\frac{1}{2}\eta_2^{\bar a})\,,\\\label{Gauss-Apm}
\triangle^{\scriptscriptstyle(N)} A_{\pm} &=
g_1^2(\eta A_{\pm}+\eta^{\bar a} C^{\bar a}_{\pm}-\frac{1}{4}\chi^{0} )\,,\\\label{Gauss-Cpm}
\triangle^{\scriptscriptstyle(N)} C^{\bar a}_{\pm}+\frac{\epsilon^{\bar{a}\bar{b}}c_{\bar{b}}}{r^2}
( C^{\scriptscriptstyle(13)}_{\pm} \pm 2iN C^2_{\pm} )
&=g_2^2(\eta C^{\bar a}_{\pm}+\eta^{\bar a} A_{\pm}-\frac{1}{4}{\chi^{\bar a}})\,,\\\label{Gauss-C2pm}
\triangle^{\scriptscriptstyle(N)}C^2_{\pm}-\frac{c_{\bar{a}}c_{\bar{a}}}{r^2}C^2_{\pm}
\pm\frac{2iN}{r^2}C^{\scriptscriptstyle(13)}_{\pm}&=g_2^2(\eta C^{2}_{\pm}\mp \frac{1}{4}\chi^2)\,,
\end{align}
\end{subequations}
where the SU(2) gauge fields have been split as $C^a=\{C^{\bar a}\,,C^2\}$, ${\bar a}=1,3$; $\epsilon^{13}=-1$,
$\epsilon^{31}=1$; $\triangle_{r}$ is the radial part of the two-dimensional Laplacian,
$\triangle^{\scriptscriptstyle(N)}=\triangle_{r}-N^2/r^2$.
Furthermore we introduce the notations $\hat a=\{0,a\}$ and $\sigma^0\equiv\mathbf{1}$,
to present more compactly the often appearing combinations
\begin{align}
\chi^{\hat a}&=\Phi_{1i}^{\star}\sigma^{\hat a}_{ij}\Phi_{j2} e^{-iN\vartheta}\,, \quad
&\eta^{\hat a}_A&=\Phi_{Ai}^{\star}\sigma^{\hat a}_{ij}\Phi_{jA},\ \hbox{no sum over $A$},\\
C^{\scriptscriptstyle(13)}&=c_3C^1-c_1C^3\,,\quad
&\eta^{\hat a}_{\pm}& =(\eta^{\hat a}_1\pm \eta^{\hat a}_2)/2\,,\notag
\end{align}
moreover to simplify the formulae somewhat we write $\eta\equiv\eta^{0}_{+}$, $\eta^{\bar a}\equiv\eta^{\bar a}_{+}$.
We have omitted the equation for $C^2_0$ from Eqs.\ \eqref{Gauss-A0}-\eqref{Gauss-C2pm},
since a straightforward application of the maximum principle leads to $C^2_0\equiv0$.
The reason behind $C^2_0\equiv0$ is the minimality of the Ansatz \eqref{minAnsatz}.
As one can see, the Gauss-type Eqs. \eqref{Gauss-A0}-\eqref{Gauss-C2pm} can be decomposed
into three equation groups, one for $\{A_0,C^{\bar a}_0\}$, and one for each $\{A_\pm,C^{\bar a}_{\pm}\}$,
 decoupled from each other.
In fact it is sufficient to consider only one of the set of eqs.\ for $\scriptstyle{\pm}$-components and impose
reality on the solutions.

From Eqs.\  \eqref{Gauss-A0}-\eqref{Gauss-C2pm} one can easily deduce the asymptotic $r\to\infty$ behaviour of
the $\omega_\alpha$ components of the gauge
potentials.
We note first that for $r\to\infty$
\begin{align}
\chi^{0}&\to0\,,\quad &\chi^{1}&\to \cos(2\alpha)\,,\quad &\chi^{2}&\to-i\,,\quad &\chi^{3}&\to -\sin(2\alpha)\,,\\
\eta^1_1&\to-\eta^1_2\to\sin(2\alpha)\,,\quad &\eta^3_1&\to-\eta^3_2\to\cos(2\alpha)\,,\quad &\eta^{\bar a}&\to 0\,,
\quad &\eta^0_A&\to 1\,,
\end{align}
then one finds that for $r\to\infty$
\begin{equation}
A_0\to\frac{1}{2}\,,\;\;A_{\pm}\to0\,,
\quad C^{\bar a}_0\to\frac{1}{2}\eta^{\bar a}_2\,,
 \quad C^{\bar a}_{\pm}\to \frac{1}{4}{\chi^{\bar a}}\,,\quad
 C^2_{\pm}\to\mp \frac{i}{4}\,.
\end{equation}
Quite interestingly, the equations for the angle-dependent components, \eqref{Gauss-Apm}-\eqref{Gauss-C2pm} can
be reduced to
a quadrature, i.e.\ to solve a single first-order, linear ordinary differential equation. The key observation
is that $A_\pm=0$ is a solution of Eq.\ \eqref{Gauss-Apm}. This is not completely obvious at first sight, since
assuming $A_\pm=0$, Eq.\ \eqref{Gauss-Apm} leads to an algebraic relation/constraint between $C^1_{\pm}$ and $C^3_{\pm}$.
A straightforward computation shows that this constraint is compatible with the remaining two coupled second-order
equations \eqref{Gauss-Cpm} and \eqref{Gauss-C2pm}. As a matter of fact one can find yet another simple algebraic
relation among
the $C^a_\pm$. In conclusion Eqs.\ \eqref{Gauss-Apm}--\eqref{Gauss-C2pm} admit a globally regular solution, which can
be given as
\begin{subequations}\label{Apm-C1pm}
\begin{align}\label{Apm}
A_\pm&=0\,,\\\label{C1pm}
r\eta^3{C^1_\pm}^{\prime}&=-(c_3\eta + N\eta^3_{-})C^1_\pm + (c_3\chi^1+i N\eta_1^3\eta_2^3/\chi^2)/4\,,
\end{align}
\end{subequations}
and in terms of the solution of Eq.\ \eqref{C1pm},
the remaining functions $C^2_\pm, C^3_\pm$ can be found from the following algebraic relations:
\begin{subequations}\label{C2-C3sol}
\begin{align}
\eta^1 C^1_{\pm}+\eta^3 C^3_{\pm}-\frac{1}{4}\chi^{0}&=0\,,\\
\chi^1C^1_{\pm} \pm \chi^2C^2_{\pm}+\chi^3C^3_{\pm}&=0\,.
\end{align}
\end{subequations}
It follows from Eqs.\ (\ref{Apm-C1pm}--\ref{C2-C3sol}) that $C^{\bar{a}}_-=C^{\bar{a}}_+$ and $C^{2}_-=-{C^2_{+}}^*$,
which implies that $C^{\bar{a}}_\pm$ are real and $C^{2}_\pm$ are imaginary.

Using the solution of the Gauss constraints \eqref{C2-C3sol}, the electric energy density, $Q$,
simplifies to
\begin{equation}\label{Q-sol}
\begin{alignedat}{3}
&Q     &&= s^2 Q_s + m^2 Q_m + ms Q_{ms}T(\vartheta)\,,\\
&Q_s   &&= \left[ \eta^0_2 (1-A_0) - C^{\bar{a}}_0\eta^{\bar{a}}_2 \right]\,,\\
&Q_m   &&= \left[ 2 C^1_+ \left( \frac{\eta^1\chi^3}{\eta_3} -\chi^1 \right) + \frac{\eta^0_-\eta^3_-}{2\eta^3}\right]\,,\\
&Q_{ms} &&= \left[ -A_0 \chi^0 - C_0^{\bar{a}} \chi^{\bar{a}} + 2C^1_+ \left( \frac{\eta^1\eta^3_2}{\eta^3} -\eta^1_2\right) +\frac{\eta^3_1 \chi^0}{2\eta_3}\right]\,,\\
\end{alignedat}
\end{equation}
where $T(\vartheta)=\cos(N\vartheta+\mu)$.
In the angular momentum density  $J$ [Eq.\ \eqref{eq:T02FMred}], $\tilde{Q}$ takes the form
\begin{equation}\label{angmom-sol}
\begin{alignedat}{2}
&\tilde{Q}          &&= sNQ_s +  m N J_m T(\vartheta)\,,\\
&J_m                 &&=-\frac{1}{2\eta^3} \left[ 4i\chi^0\chi^2 C^1_+ - \chi^0\eta^3_- \right]\,.
\end{alignedat}
\end{equation}
The total electric energy, and the longitudinal and the angular momenta are given as
\begin{equation}
  \label{eq:Qtot}
  E_0 = \frac{\omega_0^2+\omega_3^2}{4}(s^2 Q_s^{\rm tot} + m^2 Q_m^{\rm tot})\,,\quad
  P=\frac{\omega_0\omega_3}{2}(s^2 Q_s^{\rm tot} + m^2 Q_m^{\rm tot})\,,\quad
  J^{\rm tot}= \omega_0sNQ_s^{\rm tot}\,,
\end{equation}
where $Q_{m,s}^{\rm tot} = \int d^2 x Q_{m,s}$.

\begin{figure}
\noindent\hfil\includegraphics[scale=.5,angle=-90]{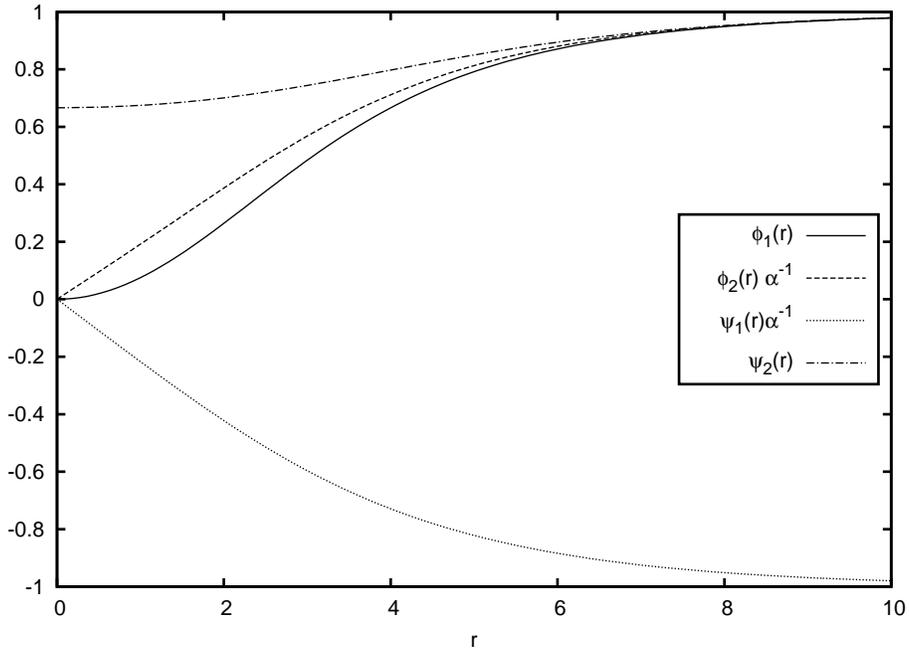}
\caption{Scalar field profiles of a coincident composite vortex for $g_1=0.4$, $g_2=1$, $\alpha=0.05$}
\label{fig:bgrscalars}
\end{figure}

\begin{figure}
\noindent\hfil\includegraphics[scale=.5,angle=-90]{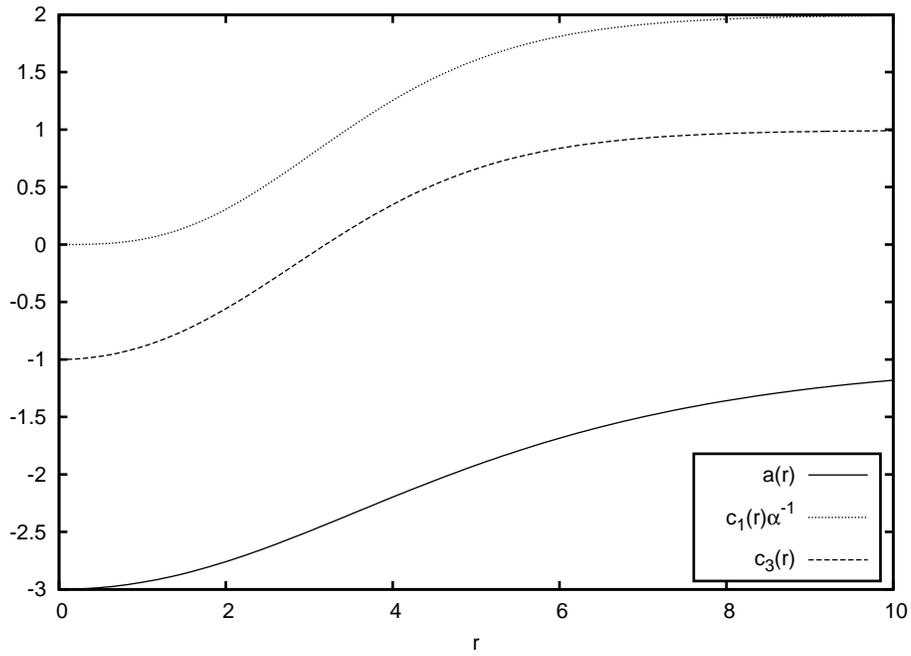}
\caption{Gauge field profiles of a coincident composite vortex for $g_1=0.4$, $g_2=1$, $\alpha=0.05$}
\label{fig:bgrgauge}
\end{figure}

\begin{figure}
\noindent\hfil\includegraphics[scale=.5,angle=-90]{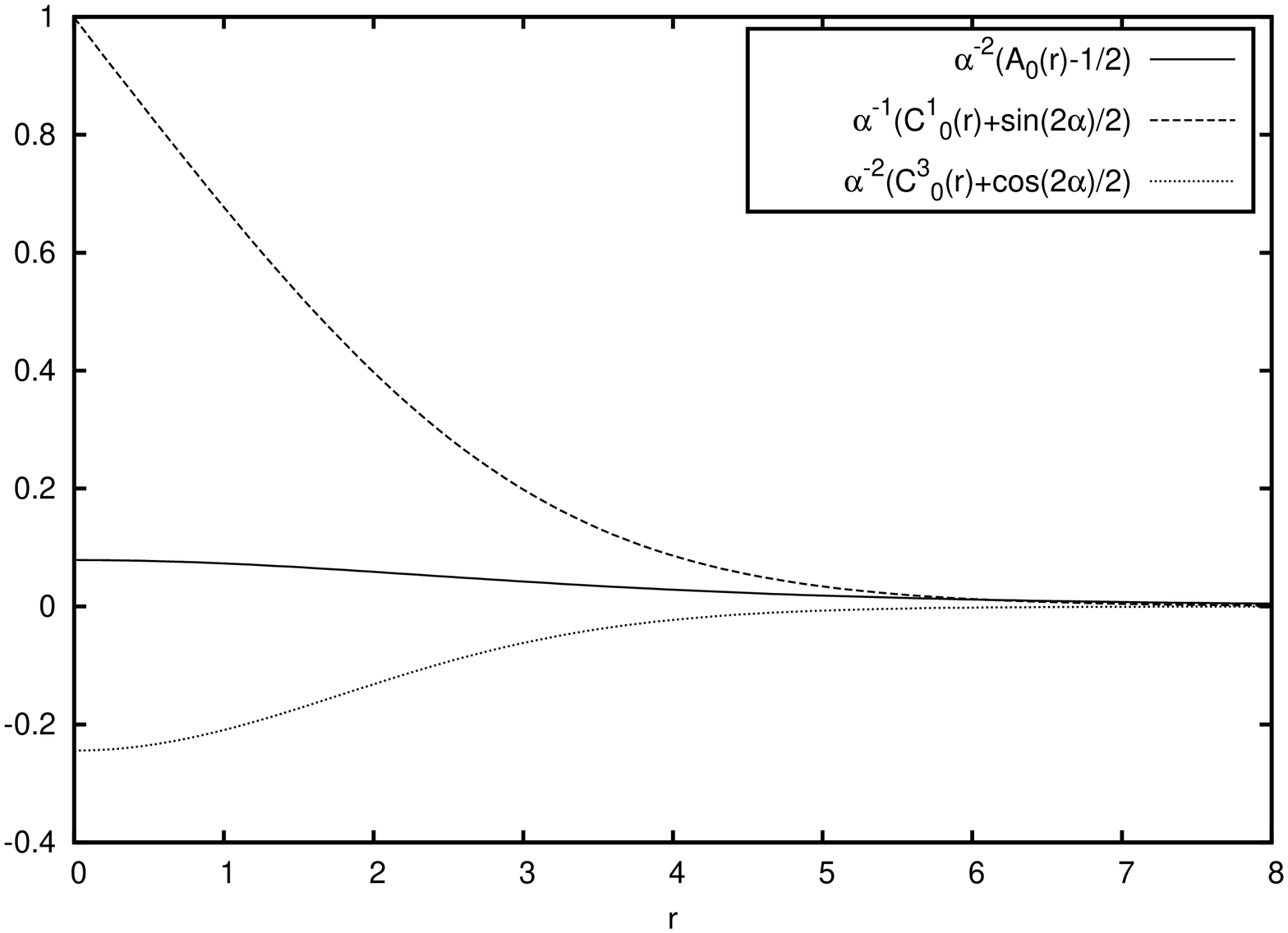}
\caption{Out-of-plane gauge field components, $A_0$, $C^1_0$, $C^3_0$, for $g_1=0.4$, $g_2=1$, $\alpha=0.0.5$}
\label{fig:s0s3a}
\end{figure}

\begin{figure}
\noindent\hfil\includegraphics[scale=.5,angle=-90]{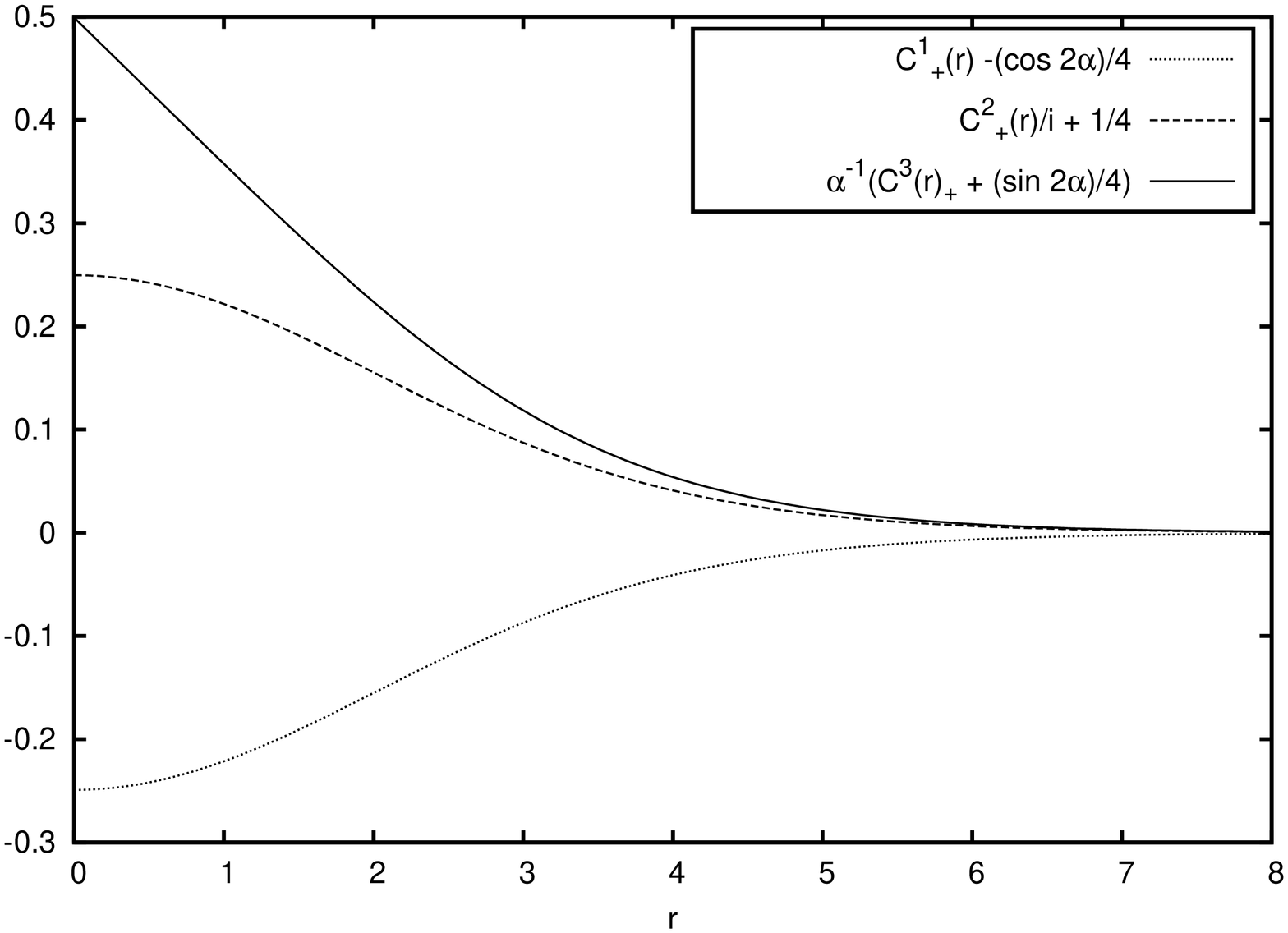}
\caption{Out-of-plane gauge field components, $C^1_+$, $C^2_+$, $C^3_+$ for $g_1=0.4$, $g_2=1$, $\alpha=0.05$}
\label{fig:s1a}
\end{figure}

\begin{figure}[t]
\noindent\hfil\includegraphics[scale=.5,angle=-90]{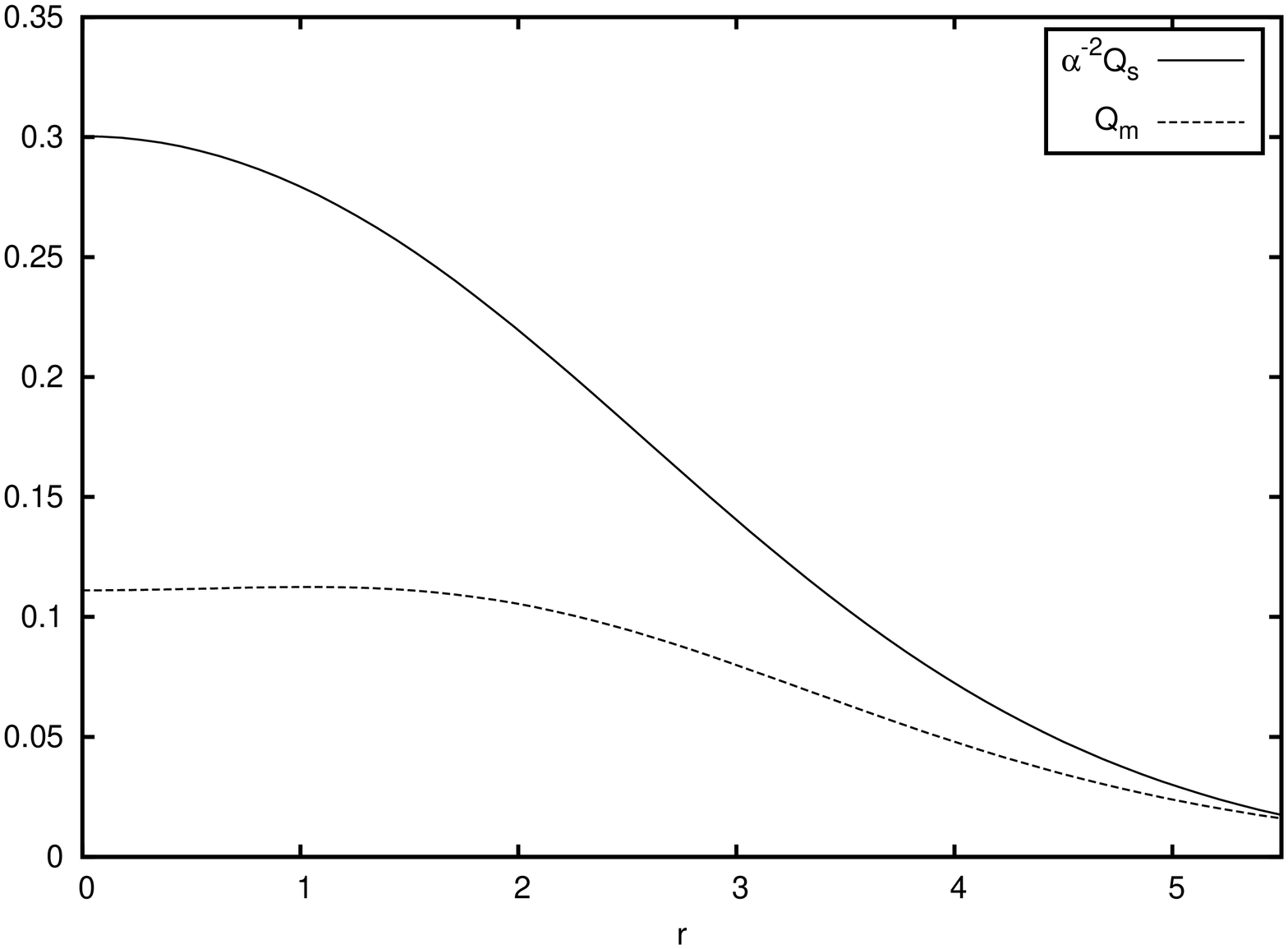}
\caption{The electric energy density terms $Q_s$, $Q_m$, for $g_1=0.4$, $g_2=1$, $\alpha=0.05$}
\label{fig:s0s3c}
\end{figure}

We have plotted the radial components of twisted coincident vortices for $N=-1$. In Figs.\ \ref{fig:bgrscalars} and \ref{fig:bgrgauge}, the planar
components are displayed (the background, planar solution). In Figs.\ \ref{fig:s0s3a}, resp.\
\ref{fig:s1a}, the Fourier components of the out-of-plane gauge fields are shown. The charge density terms $Q_s$ and $Q_m$ are plotted in Fig.\ \ref{fig:s0s3c}. Their integrals over the plane are given in Table \ref{tab:ergs}.

\begin{table}
\begin{center}
\begin{tabular}{|c |c||c| c|}
\hline
$g_1$ &  $\alpha$ & $Q_s^{\rm tot}$ & $Q_m^{\rm tot}$ \\
\hline\hline
0.4  & 0       & 0         & 6.283 \\
     & 0.05    & 0.0249    & 6.222 \\
     &0.785398 & 3.118     & 3.591 \\
\hline
0.77 & 0       & 0         & 6.507 \\
     & 0.05    & 0.0243    & 6.261 \\
     &0.785398 & 2.966     & 3.591 \\
\hline
2.33 & 0       & 0         & 6.222 \\
     & 0.05    & 0.0232    & 6.211 \\
     &0.785398 & 2.683     & 3.586 \\
\hline
\end{tabular}
\end{center}
\caption{$Q_{m}^{\rm tot}$, $Q_{s}^{\rm tot}$ for $g_2=1$  and different values of $g_1,\alpha$.
Note that the magnetic energy is $E_1=4\pi$ and $0.785398 \approx \pi/4$.}
\label{tab:ergs}
\end{table}

In Table \ref{tab:ergs}, there is a striking difference between the energy of a vortex string with $s\ne 0$, $m=0$
and $s=0$, $m\ne 0$, although they are counterparts in the sense that they have the same frequency.
Noting that the vortex with $m=0$ is rotationally symmetric, while the one with $m\ne 0$ is not, only the magnitude of
the difference is surprising.

For an explanation of the magnitude of the above-mentioned energy difference,
let us apply perturbation theory, expanding the solution in powers of $\alpha$ [see Eq.\ (\ref{red_vac})], assuming $\alpha\ll 1$.
An expansion of the background vortex as
\begin{equation}
  \label{eq:BGpertS}
  \begin{aligned}
    \phi_1 &= \phi_1^{(0)} + \alpha^2 \phi_1^{(2)}+\dots\,,\\
    \phi_2 &= \alpha \phi_2^{(1)}+\dots\,,
  \end{aligned}\quad\quad
  \begin{aligned}
    \psi_1 &= \alpha \psi_1^{(1)}+\dots\,,\\
    \psi_2 &= \psi_2^{(0)} + \alpha^2 \psi_2^{(2)}+\dots\,,\\
  \end{aligned}\quad\quad
  \begin{aligned}
    a \, &= a^{(0)}  + \alpha^2 a^{(2)}+\dots\,,\\
    c_3  &= c_3^{(0)}+ \alpha^2 c_3^{(2)}+\dots\,,\\
    c_1 &= \alpha c_1^{(1)} + \dots
  \end{aligned}
\end{equation}
can be substituted into Eqs.\ (\ref{eq:Bogeqs2}-g), yielding a consistent solution.
Note, that in the $\alpha^0$ order, the vortex is always gauge equivalent to diagonal one.
If $\alpha=\pi/2$, the configuration can also be brought to a diagonal form \cite{Auzzi}.
The field components $A_0, A_{\pm}, C^{\bar{a}}_0, C^a_\pm$ are expanded as
\be\label{eq:OOPexpand}
\begin{aligned}
A_0 &= \frac{1}{2} + \alpha^2 A_0^{(2)} \,,\\
C^1_0 &= \alpha C^{1(1)}_0\,,\\
C^3_0 &= -\frac{1}{2} + \alpha^2 C^{3(2)}_0\,,\\
\end{aligned}\quad\quad
\begin{aligned}
C^1_\pm &= C^{1(0)}_\pm + \alpha^2 C^{1(2)}_\pm \,,\\
C^{2}_\pm &= C^{2(0)}_\pm + \alpha^2 C^{2(2)}_\pm\,,\\
C^3_\pm &= \alpha C^{3(1)}_\pm\,,
\end{aligned}
\ee
and substituting into Eq.\ (\ref{Q-sol}) yields
\be\label{eq:QsQmexpand}
\begin{aligned}
Q_s &= \alpha^2 \left[ (\psi_1^{(1)})^2 -2 C_0^{1(1)} \psi_1^{(1)} - (A_0^{(2)} - C_0^{3(2)})(\psi_2^{(0)})^2 \right] + \dots\,,\\
Q_m &= - 2 C^{1(0)}_+ \phi_1^{(0)}\psi_2^{(0)} + \frac{1}{2}\left[ (\phi_1^{(0)})^2 + (\psi_2^{(0)})^2 \right]+\dots\,.
\end{aligned}
\ee
The two orders of $\alpha$ between the leading terms of $Q_s$ and $Q_m$ in Eq.\ (\ref{eq:QsQmexpand}) explain the
magnitude of the energy difference between the same planar vortex twisted with the same frequency,
either with a diagonal or with an off-diagonal twisting matrix.

Finally, we give some arguments for the stability. The conserved charge $Q$ is strongly localised, and therefore small perturbations
cannot change its value. Planar vortices, being absolute minima of the energy in
their topological sector, are stable. If there were an instability, it would be expected to manifest itself as an 
energy-reducing deformation along the $z$ axis. In the case of the twisted semilocal vortices of Ref.\ \cite{FRV}, with 
$\omega_\alpha$ timelike, such deformations indeed exist \cite{FL}, however, they correspond to the same type of 
instability as those of ANO vortices embedded in a two-component extended Abelian Higgs model \cite{Hindmarsh}. 
In the present case, however, such potential instabilities are absent.  
The spectrum of the perturbation modes of the untwisted vortices are gapped, moreover the planar and 
off-planar perturbation modes decouple. Therefore at least for small values of the twist, it cannot change the sign of 
the otherwise non-vanishing positive eigenvalues.

\section{Conclusions}\label{sec:sum}
In this paper, we have constructed charged, stationary rotating non-Abelian vortex strings in a $U(2)_{\rm gauge}\times SU(2)_{\rm flavor}$ theory.
The scalar fields rotate around the string axis, and they have a (matrix) phase depending linearly on $x^\alpha=(t,z)$ as
\[
\Phi(x^\mu) = \Phi(x^i)\exp\left(\frac{i}{2}M\omega_\alpha x^\alpha\right)\,,
\]
which is referred to as a twist.
We considered here the case $\omega_\alpha\omega^\alpha =0$, in which, the planar equations decouple from those of the $t,z$ components.  The energy
contribution due to the twist, and the $z$ component of the momentum are both proportional to the Noether charge corresponding to the
flavor symmetry generated by the matrix $M$.

Adding twist to the coincident composite vortices of Ref.\ \cite{Auzzi}, leads to some striking phenomena.
These vortex strings carry total angular momentum, unless $M$ is purely off-diagonal. If $M$ is non-diagonal,
the vortex strings are not rotationally symmetric, although, all their planar cross sections are. This
is explained by the fact, that the nontrivial realizations of rotations and $z$-translations
act on them non-commutatively. The energy of a solution, which breaks rotational symmetry, is
significantly larger than that of its rotationally symmetric counterpart.

The analysis of vortex hair for charged rotating asymptotically anti-de Sitter black holes has
revealed interesting features particularly in the holographic context of
the gauge/gravity duality \cite{szoeros}. The interplay between the angular momentum of vortices such as those
constructed here and the black hole angular momentum could give rise to interesting effects. We hope to discuss
this issue in a forthcoming work.

\section*{Acknowledgements} This work has been supported by the grants OTKA K101709 and TÉT 10-1-2011-0071
and MINCYT-Argentina HU/10/01. F.A.S.\ is partially financed by PIP-CONICET, PICT-ANPCyT and CICBA. Á.L.\ is partially financed
by The Hellenic Ministry of Education: Education and Lifelong Learning Affairs, and European Social Fund: NSRF 2007-2013,
Aristeia (Excellence) II (TS-3647).

\appendix
\section{Notation and conventions}\label{app:conventions}
Here we summarize the definition of the gauge-field strength tensor, covariant derivatives, etc.
The signature of the flat Minkowskian metric used here is $(+,-,-,-)$,
\begin{align}
F_{\mu\nu}&=\partial_\mu{A}_\nu-\partial_\nu{A}_\mu\,,\nonumber \\
G^a_{\mu\nu}&=\partial_\mu C^a_\nu -\partial_\nu C^a_\mu
+\epsilon^{abc}C^b_\mu{C}^c_\nu\,,\nonumber \\
\nonumber  \\
D_\mu\Phi_A &=\left(\partial_\mu-\frac{i}{2}\,A_\mu
-\frac{i}{2}\,\sigma^a C^a_\mu\right)\Phi_A\,,
 \label{defs1}
\end{align}
with $\sigma^a$, ($a=1,2,3$) denoting the Pauli matrices,
$\sigma^a\sigma^b=\delta^{ab}+i\epsilon^{abc}\sigma^c$.
For later use,  the covariant derivative of adjoint representation fields,
$\hat{D}_\mu \Sigma^a = \partial_\mu \Sigma^a +\varepsilon_{abc}C_\mu^b\Sigma^c$.

The Yang-Mills-Higgs equations are
\begin{equation}\label{eq:YM}
  \begin{aligned}
    \partial^\mu F_{\mu\nu} &= g_1^2 J^0_\nu\,,\\
    {\hat{D}}^\mu G_{\mu\nu}^a &= g_2^2 J_\nu^a\,,
  \end{aligned}
\end{equation}
where the color currents are
\begin{equation}\label{eq:ccurr}
\begin{aligned}
J_\mu^0 &= \frac{i}{2}\Tr \left[ (D_\mu\Phi)^\dagger \Phi - \Phi^\dagger D_\mu\Phi\right]\,,\\
J_\mu^a &= \frac{i}{2}\Tr \left[ (D_\mu\Phi)^\dagger \sigma^a \Phi - \Phi^\dagger\sigma^a D_\mu\Phi\right]\,.
\end{aligned}
\end{equation}

The flavor current, i.e., the Noether current corresponding to the global $SU(2)$ flavor symmetry is
\begin{equation}
  \label{eq:flcurr}
  K^{\hat{a}}_\mu = \frac{i}{2}\Tr \left[ D_\mu \Phi \sigma^{\hat{a}} \Phi^\dagger - \Phi \sigma^{\hat{a}} D_\mu\Phi^\dagger \right]\,,
\end{equation}
where the component $K_\mu^0=-J_\mu^0$ has been introduced for the sake of convenience; a $U(1)$ transformation agrees
with a gauge transformation with a constant (global) phase.

\section{Numerical data}\label{app:numdata}
In this appendix, we present some numerical data of the untwisted vortices (Table \ref{tab:bgrnum}) and of
the twisted strings (Table \ref{tab:gaussnum}). The shooting parameters in the Tables are defined at the origin as
\begin{equation}
  \label{eq:bgr_ser}
  \begin{aligned}
    \phi_1(r) &= f_1 r^2 + {\cal O}(r^4)\,,\\
    \phi_2(r) &= f_2r + {\cal O}(r^3)\,,
  \end{aligned}\quad
  \begin{aligned}
    \psi_1(r) &= p_1 r + {\cal O}(r^3)\,,\\
    \psi_2(r) &= p_2 + {\cal O}(r^2)\,,
  \end{aligned}
\end{equation}
and similarly
\begin{equation}
  \label{eq:gauss_ser}
  \begin{aligned}
    A_0   &= s_{00} + {\cal O}(r^2)\,,\\
    C^1_0 &= s_{10} r + {\cal O}(r^3)\,,\\
    C^3_0 &= s_{30} + {\cal O}(r^2)\,,
  \end{aligned}\quad
  \begin{aligned}
    C_p\, &= s_pr^2 + {\cal O}(r^4)\,,\\
    C_m\, &= s_m + {\cal O}(r^2)\,,\\
    C^3_+ &= s_3r + {\cal O}(r^3)\,.
  \end{aligned}
\end{equation}
where $C_p = C^1_+ + i C^2_+$ and $C_m = C^1_+ - i C^2_+$.\\
We recall that $A_\pm=0$, $C^2_0=0$, $C^1_+=C^1_-$, $C^3_+=C^3_-$, $C^2_+=-C^2_-$ and that for $\alpha=0$,
$s_{00} = s_{10}=s_{30} = s_{m} = s_{3} = 0$. The relations (\ref{C2-C3sol}) give
\begin{subequations}\label{eq:C2-C3sol-sh}
\begin{align}
-p_1 s_m + p_2 s_3 + f_2& =0\,,\\
f_1s_m - f_2 s_3&=0\,.
\end{align}
\end{subequations}
Equations (\ref{eq:C2-C3sol-sh}) are trivially satisfied for $\alpha=0$. For nonzero values of $\alpha$,
the numerical errors in Eqs.\ (\ref{eq:C2-C3sol-sh}) vary between $10^{-11}$-- $10^{-8}$. Of the values considered,
the minimal error occurs  for $g_2=2.33$, $\alpha=0.05$, when Eq.\ (\ref{eq:C2-C3sol-sh}b) is satisfied to a precision of $3\times 10^{-12}$,
while the maximal error occurs for $g_2=2.33$, $\alpha=0.785398$, when Eq.\ (\ref{eq:C2-C3sol-sh}a) is satisfied to a precision of $4\times10^{-8}$. Over the
intervals shown in the figures, the errors in the algebraic constraints (\ref{C2-C3sol}a-b) remain below $6\times 10^{-5}$.

\begin{table}
\begin{center}
\begin{tabular}{|c|c||c|c|c|c|}
\hline
$g_1$ &  $\alpha$ & $f_1$ & $f_2$ & $p_1$ & $p_2$ \\
\hline\hline
0.4   & 0        & 0.07863    & 0          & 0          & 0.6672 \\
      & 0.05     & 0.07855    & 0.009484   & -0.01094   & 0.6666 \\
      & 0.785398 & 0.05975    & 0.1485     & -0.1617    & 0.5176 \\
\hline
0.77  & 0        & 0.1786     & 0          & 0          & 0.9002 \\
      & 0.05     & 0.1785     & 0.01635    & -0.01908   & 0.8996 \\
      & 0.785398 & 0.1435     & 0.2665     & -0.2930    & 0.7188 \\
\hline
2.33  & 0        & 0.4397     & 0          & 0          & 1.2724 \\
      & 0.05     & 0.4397     & 0.02963    & -0.03542   & 1.2720 \\
      & 0.785398 & 0.4350     & 0.5589     & -0.6301    & 1.1314 \\
\hline
\end{tabular}
\end{center}
\caption{Shooting parameters of planar vortices for $g_2=1$  and different values of $g_1,\alpha$}
\label{tab:bgrnum}
\end{table}

\begin{table}
\begin{center}
\begin{tabular}{|c|c||c|c|c||c|c|c|}
\hline
$g_1$ &  $\alpha$ &  $s_{00}$  & $s_{10}$ & $s_{30}$  & $s_m$ & $s_p$ & $s_{3}$\\
\hline\hline
0.4 & 0           &                &              & 		  &            & 0.05893   &                \\
    & 0.05        & 0.5002         & -0.01637     & -0.4981       & -0.0008571 & 0.05881   & -0.007099       \\
    & 0.785398    & 0.5250         & -0.2231      & -0.08282      & -0.2007    & 0.03250   & -0.08075        \\
\hline
0.77& 0	          &		   &              &               &            & 0.09920   & \\
    & 0.05        & 0.5005         & -0.02117     & -0.4980       & -0.0008310 & 0.09901   & -0.009071 \\
    & 0.785398    & 0.5682         & -0.2231      & -0.08282      & -0.1960    & 0.05680   & -0.1055 \\
\hline
2.33& 0           &                &              &               &            & 0.1728    & \\
    & 0.05        & 0.5016         & -0.02777     & -0.4977       & -0.0007835 & 0.1725    & -0.01163 \\
    & 0.785398    & 0.7470         & -0.2926      & -0.06037      & -0.1850    & 0.1121    & -0.1440 \\
\hline
\end{tabular}
\end{center}
\caption{Shooting parameters of twisted strings for $g_2=1$  and different values of $g_1,\alpha$}
\label{tab:gaussnum}
\end{table}

\begin{thebibliography}{99}
\def\refttl#1{{\sl ``#1''}, }%
%
\bibitem{ANO}
A.A.~Abrikosov, \refttl{On the Magnetic Properties of Superconductors of the Second Group}
{\sl Sov.\ Phys.\ JETP}, {\bf 5} (1957) 1174;
H.B.~Nielsen and P.~Olesen, \refttl{Vortex-line models for dual strings}
{\sl Nucl.\ Phys.}, {\bf B 61} (1973) 45.
%
\bibitem{bog}
E.~B.~Bogomol'nyi, \refttl{The stability of classical solutions} {\sl Sov.\ J.\ Nucl.\ Phys.} {\bf 24}, 449, (1976).
%
\bibitem{deVegaSchap1}
H.~J.~de Vega and F.~A.~Schaposnik, \refttl{Classical vortex solution of the Abelian Higgs model} {\sl Phys.\ Rev.} {\bf D14}, 1100, (1976).
%
\bibitem{fidel} H.~J.~de Vega and F.~A.~Schaposnik, \refttl{Electrically charged vortices in non-Abelian gauge theories with Chern--Simons term} {\sl Phys.\ Rev.\ Lett.} {\bf 56}, 2564 (1986);
H.~J.~de Vega and F.~A.~Schaposnik, \refttl{Vortices and electrically charged vortices in non-Abelian gauge theories}
{\sl Phys.\ Rev.} {\bf D34}, 3206--3213, (1986).

\bibitem{nav1}
A.~Hanany and D.~Tong, \refttl{Vortices, instantons and branes} {\sl JHEP} {\bf 0307}, 037 (2003) \arxiv{hep-th/0306150};\\
R.~Auzzi, S.~Bolognesi, J.~Evslin, K.~Konishi and A.~Yung, \refttl{Nonabelian superconductors: Vortices and confinement in N = 2 SQCD} {\sl Nucl.\ Phys.} {\bf B 673}, 187 (2003) \arxiv{hep-th/0307287};\\
N.~Markov, A.~Marshakov, A.~Yung, \refttl{Non-Abelian vortices in $\mathcal{N}=1*$ gauge theory} {\sl Nucl.\ Phys.} {\bf B709}, 267 (2005) \arxiv{hep-th/0408235}.


\bibitem{Shifman} M.~Shifman and A.~Yung, {\sl Supersymmetric solitons}, CUP, 2009.
%
\bibitem{Auzzi} R.~Auzzi, M.~Shifman, A.~Yung, \refttl{Composite non-Abelian flux tubes in $\mathcal{N}=2$ sQCD}
{\sl Phys.\ Rev.} {\bf D73}, 105012 (2006) \arxiv{hep-th/0511150}.

\bibitem{Eto-review} M.~Eto, Y.~Isozumi, M.~Nitta, K.~Ohashi, and N.~Sakai,
\refttl{Solitons in the Higgs phase: the moduli matrix approach}\ {\sl J.\ Phys.\ A: Math.\ Gen.}
{\bf 39} (2006) R315--R392\arxiv{hep-th/0602170};\\
M.~Eto, Y.~Isozumi, M.~Nitta, K.~Ohashi, and N.~Sakai, \refttl{Moduli space of non-Abelian vortices}
{\sl Phys.\ Rev.\ Lett.} {\bf 96}, 161601 (2006) \arxiv{hep-th/0511088};\\
M.~Eto, K.~Konishi, G.~Marmorini, M.~Nitta, K.~Ohashi, W.~Vinci, and N.~Yokoi, \refttl{Non-Abelian vortices of higher winding number}
{\sl Phys.~Rev.} {\bf D74}, 065021 (2006)\arxiv{hep-th/0607070};\\
M.~Eto, T.~Fujimori, S.B.~Gudnason, K.~Konishi, M.~Nitta, K.~Ohashi, and W.~Vinci, \refttl{Constructing non-Abelian vortices with
arbitrary gauge groups} {\sl Phys.\ Lett.} {\bf B669}: 98--101 (2008)\arxiv[hep-th]{0802.1020}.

\bibitem{ConfMono} D.~Tong, \refttl{Monopoles in the Higgs phase} {\sl Phys.~Rev.} {\bf D69} 065003 (2004) \arxiv{hep-th/0307302};
A.~Hanany and D.~Tong, \refttl{Vortex strings and four-dimensional gauge dynamics} {\sl JHEP} {\bf 0404}, 066 (2004) \arxiv{hep-th/0403158};
R.~Auzzi, S.~Bolognesi, J.~Evslin, \refttl{Monopoles can be confined by 0,1 or 2 vortices} {\sl JHEP} {\bf 0502} 046 (2002)
\arxiv{hep-th/0411074}.

\bibitem{Gorsky} A.~Gorsky, M.~Shifman, and A.~Yung, \refttl{Non-Abelian Meissner effect in Yang-Mills
theories at weak coupling} {\sl Phys.\ Rev.} {\bf D71}, 045010 (2005) \arxiv{hep-th/0412082}.

\bibitem{Collie} B.~Collie, \refttl{Dyonic non-Abelian vortices} {\sl J.\ Phys.} {\bf A42} (2009) 085404 \arxiv[hep-th]{0809.0394}.

\bibitem{Eto-dyonic} M.~Eto, T.~Fujimori, M.~Nitta, K.~Ohashi, and N.~Sakai, \refttl{Dynamics of non-Abelian vortices}
{\sl Phys.\ Rev.} {\bf D84} (2011) 125030 \arxiv[hep-th]{1105.1547}.

\bibitem{Eto-eff} M.~Eto, Y.~Isozumi, M.~Nitta, K.~Ohashi, and N.~Sakai, \refttl{Manifestly supersymmetric effective Lagrangians
on BPS solitons} {\sl Phys.\ Rev.} {\bf D73}, 125008 (2006) \arxiv{hep-th/0602289}.

\bibitem{DW} D.~Tong, \refttl{The moduli space of BPS domain walls} {\sl Phys.\ Rev.} {\bf D66}, 025013 (2002) \arxiv{hep-th/0202012}

\bibitem{HT} A.~Hanany and D.~Tong, \refttl{On monopoles and domain walls} {\sl Commun.\ Math.\ Phys.} {\bf 266} (2006) 647-663 \arxiv{hep-th/0507140}.

\bibitem{Abraham} E.~Abraham, \refttl{Charged semilocal vortices} {\sl Nucl.\ Phys.} {\bf B399} (1993) 197-210.

\bibitem{FRV} P.~Forgács, S.~Reuillon, and M.S.~Volkov, , \refttl{Superconducting Vortices in Semilocal Models} {\sl Phys.\ Rev.\ Lett.} {\bf 96}, 041601 (2006) \arxiv{hep-th/0507246};\\
\refttl{Twisted superconducting semilocal strings} {\sl Nucl.\ Phys.}{\bf B 751} (2006) 390--418 \arxiv {hep-th/0602175};\\
Y.~Brihaye, L.~Honorez, \refttl{Twisted semilocal strings in the MSSM} {\sl Int.\ J.\ Mod.\ Phys.} {\bf A23} (2008) 581-597  \arxiv {hep-th/0701141}.
%
\bibitem{GV}
Y.~Brihaye, Y.~Verbin, \refttl{Superconducting and spinning non-Abelian flux tubes}
{\sl Phys.\ Rev.} {\bf D77} (2008) 105019  \arxiv[hep-th]{0711.1112};
J.~Garaud and M.S.~Volkov, \refttl{Superconducting non-Abelian vortices in Weinberg-Salam theory -- electroweak thunderbolts} {\sl Nucl.Phys.} {\bf B826} (2010) 174-216 \arxiv[hep-th]{0712.3589}.

\bibitem{Ferreira} L.A.~Ferreira, \refttl{Exact vortex solutions in an extended Skyrme-Faddeev model} {\sl JHEP} {\bf 0905} (2009) 001 \arxiv[hep-th]{0809.4303}.

\bibitem{FM} P.~Forgács, N.S.~Manton, \refttl{Space-Time Symmetries in Gauge Theories} {\sl Commun.\ Math.\ Phys.} {\bf 72} (1980) 15.

\bibitem{FL} J.~Garaud and M.S.~Volkov, \refttl{Stability analysis of the twisted superconducting semilocal strings}
{\sl Nucl.\ Phys.} {\bf B799} (2008) 430-455 \arxiv[hep-th]{0712.3589};
P.\ Forgács, Á.\ Lukács, \refttl{Instabilities of twisted strings} {\sl JHEP} {\bf 0912} (2009) 064 \arxiv[hep-th]{0908.2621};
B.\ Hartmann, P.\ Peter, \refttl{Can type II Semi-local cosmic strings form?} {\sl  Phys.\ Rev.} {\bf D86} (2012) 103516  \arxiv[hep-th]{1204.1270}

\bibitem{Hindmarsh} M.~Hindmarsh, \refttl{Existence and stability of semilocal strings} {\sl Phys.\ Rev.\ Lett.} {\bf 68} (1992)
1263-1266.

\bibitem{szoeros} R.~Gregory, P.C.~Gustainis, D.~Kubizňák, R.B.~Mann, and D.~Wills,
\refttl{Vortex hair on AdS black holes} {\sl JHEP} {\bf 1411} (2014) 010 \arxiv[hep-th]{1405.6507}.
\end{thebibliography}
\end{document}